# An immersive micro-manipulation system using real-time 3D imaging microscope and 3D operation interface for high-speed and accurate micro-manipulation

Kenta Yokoe, Tadayoshi Aoyama*, Toshiki Fujishiro, Masaru Takeuchi and Yasuhisa Hasegawa

**Abstract**

The use of intracytoplasmic sperm injection (ICSI), an assisted reproductive technique (ART), is increasing widely. ICSI is currently performed by specially skilled embryologists. However, with the increasing demand for ART, the shortage of skilled embryologists has become a problem. Therefore, we propose an immersive micromanipulation system that requires no special skills for efficient and accurate micromanipulation. Our proposed system is composed of a real-time three-dimensional (3D) imaging microscope and 3D operation interfaces. The 3D operation interfaces are stationary pen-type or wearable glove-type interfaces. In this system, an operator wearing a head-mounted display (HMD) and using 3D operation interfaces is immersed in a virtual micromanipulation space. The operator can move the pipettes by 3D operation interface and freely change the viewpoint. We verified that the proposed system improves the speed and accuracy of operating a pipette through two types of experiments with subjects.

## Introduction

In recent years, the use of micromanipulation to inject DNA and cells into unfertilized eggs, such as artificial insemination and transgenics, is in great demand [1, 2]. Intracytoplasmic sperm injection (ICSI) is the most widely used assisted reproductive technique (ART) globally with increasing demand annually [3, 4].

Figure 1 shows the ICSI procedure. Generally, ICSI is performed under an optical microscope using motorized micromanipulators operated by joysticks. ICSI is done by dividing the droplet into three spaces on the dish. In spaces 1 and 3, the cells before and after manipulation are placed, respectively, and in space 2, the microscopic work is performed. The ICSI operator moves one of the embryos from space 1 to space 2. When injecting sperm into the egg in space 2, the injection pipette should be inserted while avoiding the spindle. As the spindle is invisible, the operator must estimate the position of the spindle from the position of the polar body. The positions of the polar body and spindle in the embryo are shown in Fig. 2. The operator moves the embryo from space 2 to space 3 after injection. ICSI, the aforedescribed process, requires high-speed and high-precision micromanipulation [5]. Particularly, the movement of the embryo between spaces must be performed quickly and the manipulation of the embryo in space 2 must be performed accurately. Therefore, embryologists who perform ICSI require advanced skills. There is a positive correlation between the skill of the embryologist and the rate of embryo development [6].

An embryologist must be skilled to use optical microscopes and interface for moving micromanipulators. As optical microscopes capture 2-dimensional (2D) moving images from only one direction, visual information in the depth direction is limited; thus, operators need skills in recognizing depth information from 2D images. In addition, optical microscopes require complicated operations for changing a viewpoint because magnification and light intensity are also changed when the viewpoint is changed. Furthermore, the micromanipulator has 3-degrees-of-freedoms (DoFs) in translation; however, the joystick operation is a combination of 2-DoFs in translation and 1-DoF in rotation. Therefore, advanced techniques are required to freely operate the micromanipulators.

*Correspondence: tadayoshi.aoyama@mae.nagoya-u.ac.jp
Department of Micro-Nano Mechanical Science and Engineering, Nagoya University, 1 Furo-cho, Chikusa-ku, Nagoya, Aichi 464-8603, Japan

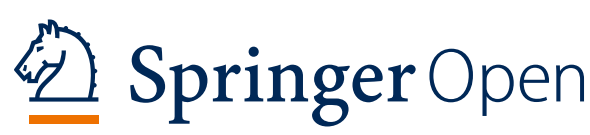

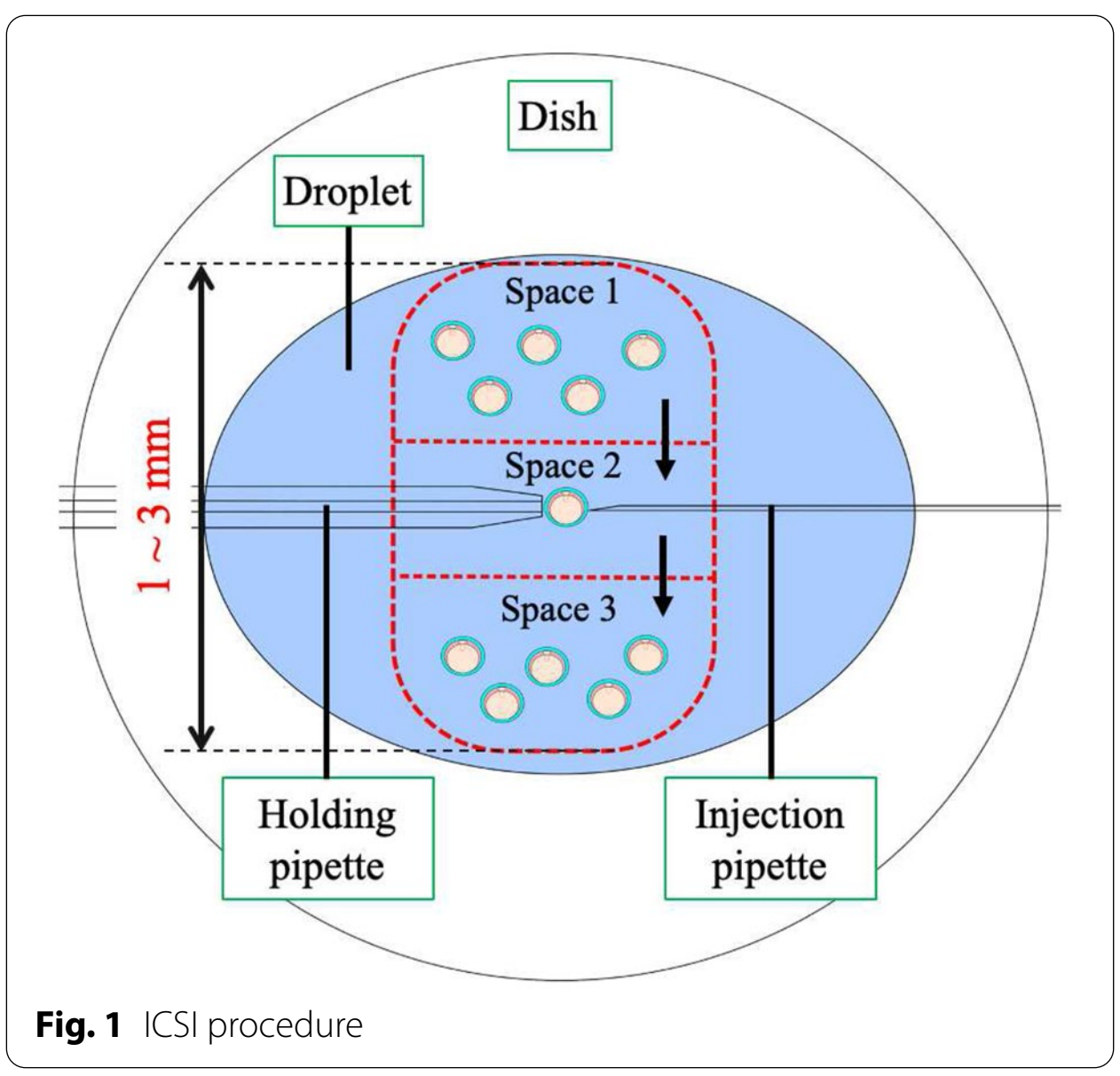

**Fig. 1** ICSI procedure

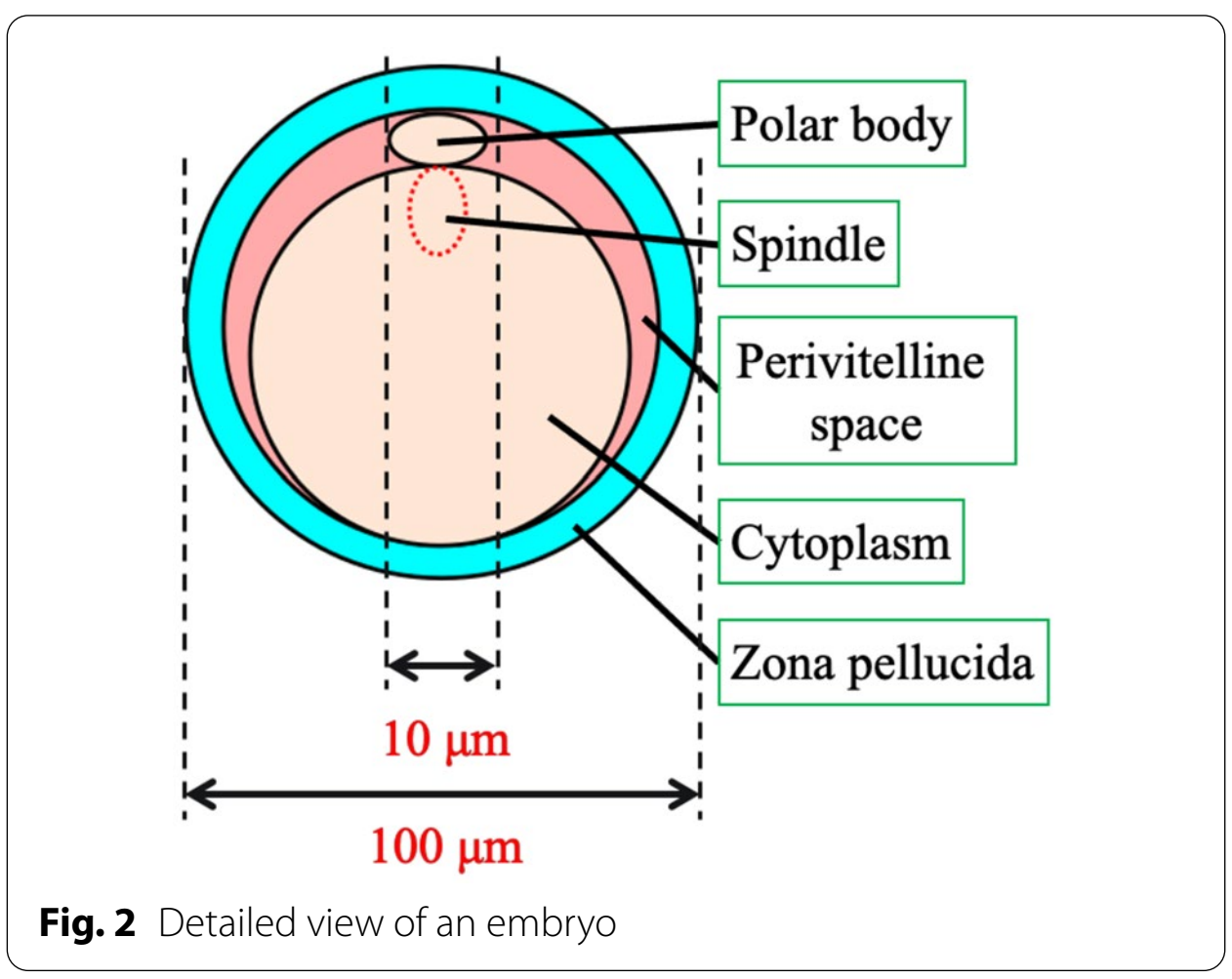

**Fig. 2** Detailed view of an embryo

Many studies have attempted to improve the visibility of microscopes. The digital holographic microscope can obtain 3D information of microscopic objects [7, 8]. However, it requires time for 3D image reconstruction and therefore is not suitable for ICSI as ICSI requires precise 3D information in real-time. Aoyama et al. presented a head-mounted display (HMD)-based microscopic imaging system [9]. Although the system enables changing the microscope's viewpoint and magnification easily, the images produced are 2D with limited depth visibility. Fujishiro et al. proposed a real-time 3D imaging microscope using focal length adjustment [10]. The system enables real-time 3D image presentation of micromanipulation; however, the viewpoint cannot be freely changed.

3D operation interfaces for micromanipulators have also been proposed. Ando et al. developed a parallel link micromanipulator operated by a dedicated 3D control device [11]. The micromanipulators used in ICSI have different shapes from the micromanipulators used in this system and therefore cannot be applied to ICSI. Further, this system has poor depth visibility; therefore, it offers no significant advantage to operating in 3D. Ammi et al. proposed a method of operating a micromanipulator using a 3D haptic device [12]. However, this system is designed for a static environment and is not suitable for ICSI, which is a dynamic environment. Onda et al. presented a system to operate optical tweezers with 3D haptic devices [13]. Embryos (100 µm) are too large to manipulate using optical tweezers because they can only exert a small force. All of these 3D operation systems were incompatible with accurate real-time 3D microscopic images, and their effectiveness is still unverified by subject experiments.

Immersive teleoperation systems using VR have been developed for many robots in recent years. Various tasks can be performed using immersive teleoperation systems in macro-spaces, such as bottle picking [14], assembly [15], and cleaning [16]. However, no immersive teleoperation system has been developed for micro-spaces. In this study, we developed an immersive micromanipulation system that enables high-speed and high-precision micromanipulation using our 3D imaging microscope [10] and 3D operation interfaces. Figure 3 shows the concept of the proposed system. The proposed system constructs an immersive micromanipulation space using the 3D imaging microscope, and the operator is immersed in the virtual manipulation space and operates the micromanipulators using 3D operation interfaces. Furthermore, the proposed system allows users to freely change their viewpoints. We verified that the proposed system improves the speed and accuracy of micromanipulation through two types of subject experiments by operating a micromanipulator.

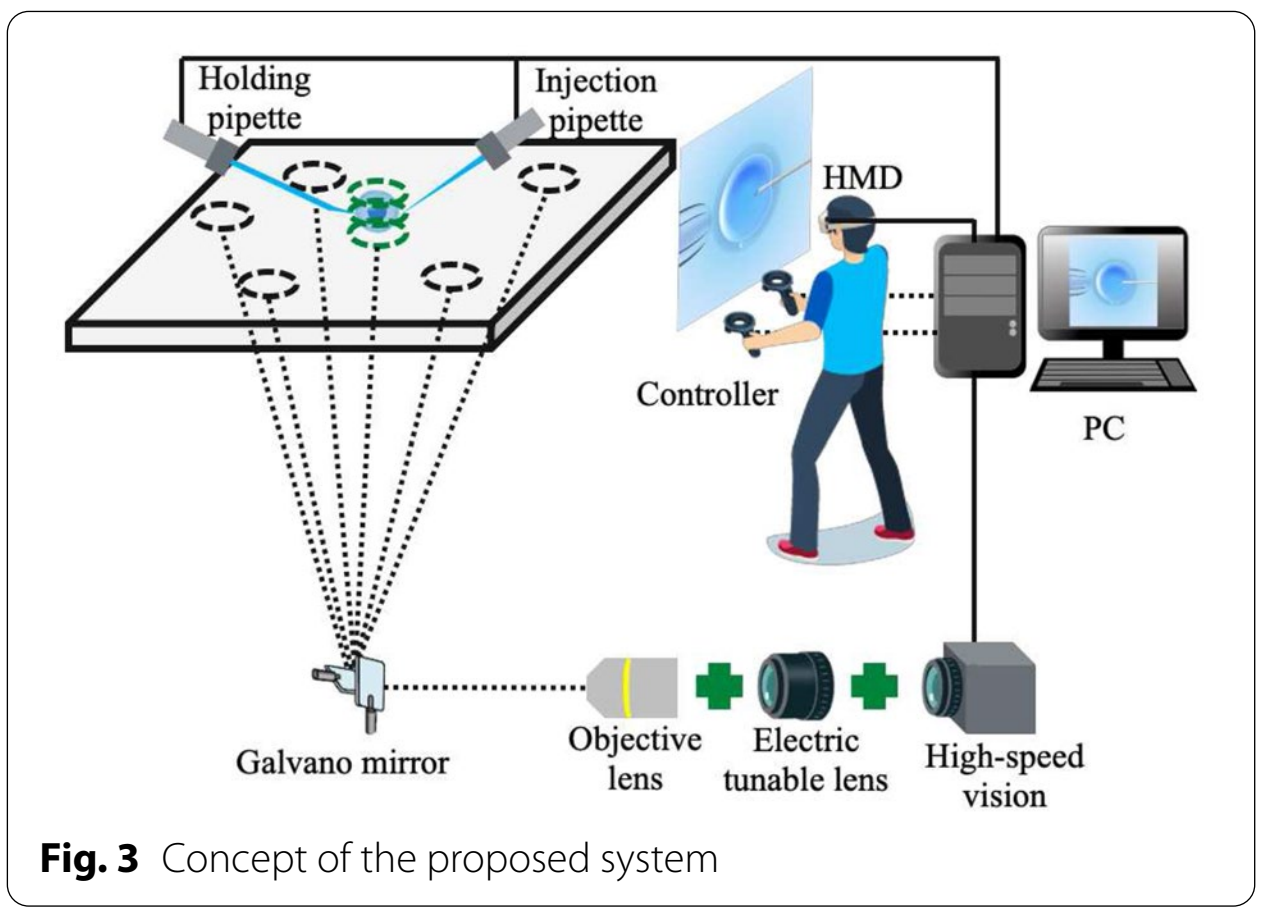

**Fig. 3** Concept of the proposed system

## Immersive micromanipulation system

### System outline

Figures 4 and 5 show the configuration and overview of the proposed immersive micromanipulation system, respectively. The system comprises an inverted microscope (IX73, OLYMPUS), an objective lens (LWD95mm, 10X, Mitutoyo), a high-speed vision system (MQ003MG-CM, Ximea), an electrically tunable lens (ETL) (EL-10-30-C-VIS-LD-MV, Optotune), a lens driver (Lens Driver 4, Optotune), a two-axis galvanometer mirror (6210HSM 6mm 532nm, Cambridge Technology), a control PC (OS Windows 10 Home 64-bit, CPU Intel(R) Core(TM) i9-9900KF 3.60-GHz, 32-GB RAM, GPU NVIDIA GeForce RTX 2080 SUPER), D/A board (PCX-340416, Interface), a light source unit (LA-HDF158AS, Hayashi Repic Co., Ltd.), micromanipulators with a joystick (TransferMan 4r, Eppendorf), two microinjectors (FemtoJet 4i and CellTram 4r Air, Eppendorf), an injection pipette (Femtotip II, Eppendorf), a holding pipette (VacuTip I, Eppendorf), a HMD (HTC VIVE Pro Eye, HTC Corporation), a wearable glove-type interface (MANUS Prime Haptic, MANUS VR Company), and a stationary pen-type interface (PHANToM Premium1.5HF, 3D Systems). The inner diameter of the injection pipette is 0.5 μm. The inner and outer diameters of the holding pipettes are 15 μm and 100 μm, respectively.

This system measures the 3D positions of the micromanipulators and the manipulated object using the focal-point adjustment function of an ETL to construct an immersive space. The operator is immersed in the immersive space and operates the micromanipulator using 3D operation interfaces. Additional file 1 shows

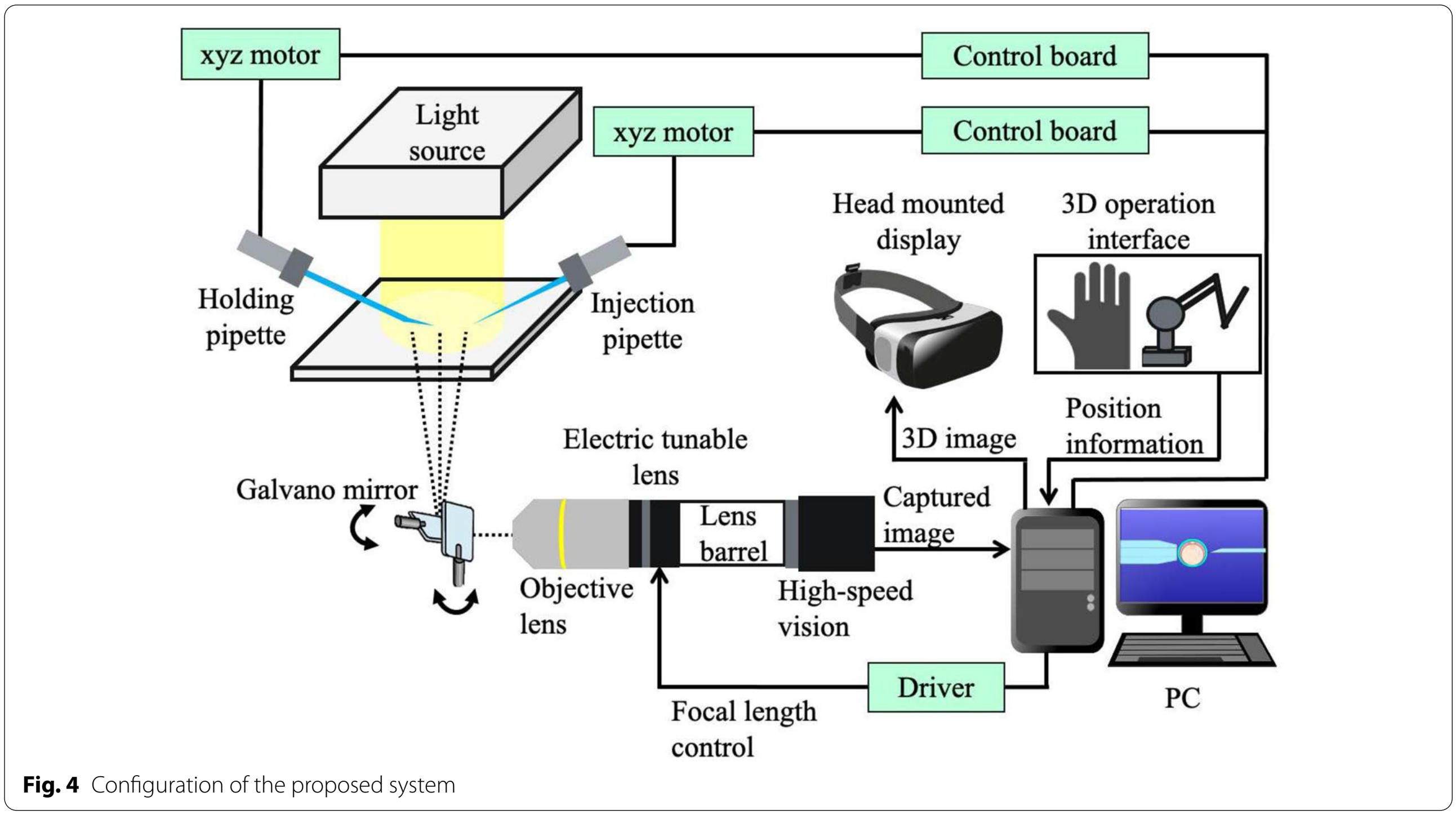

**Fig. 4** Configuration of the proposed system

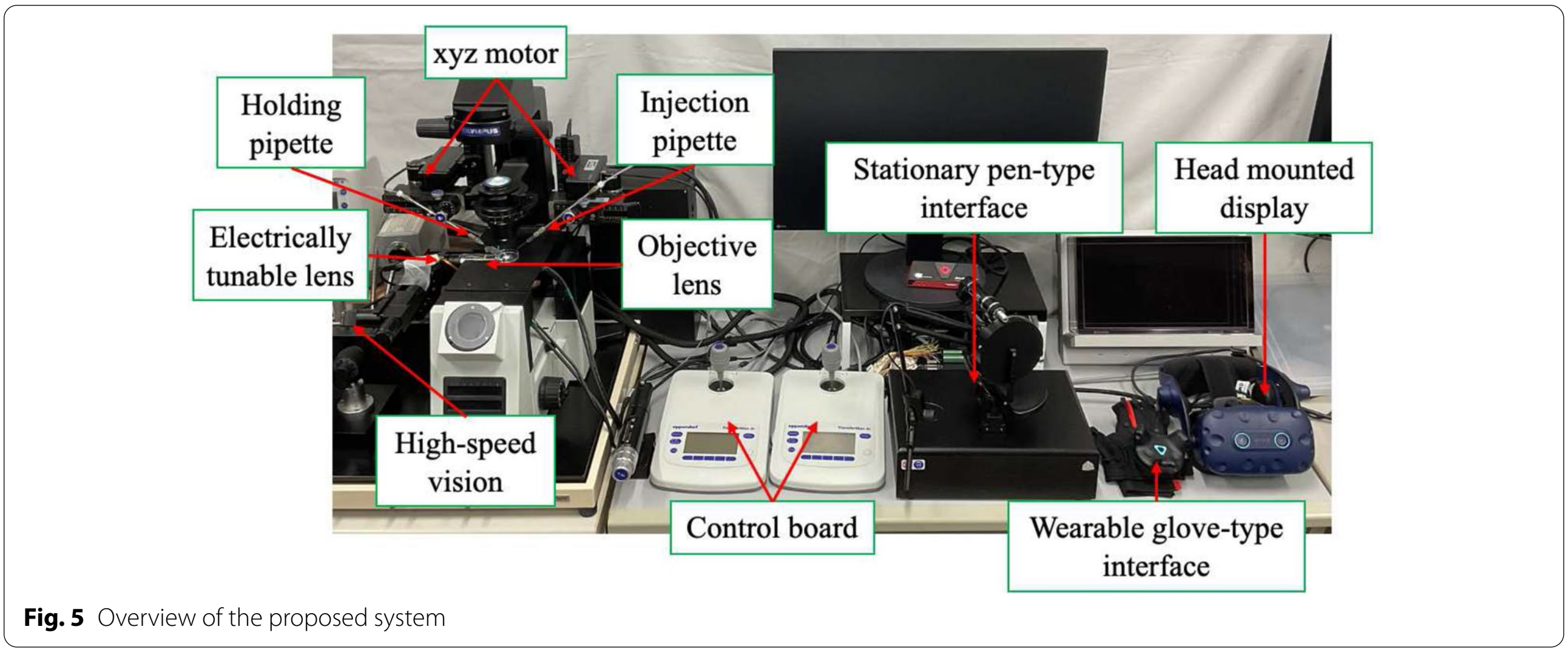

**Fig. 5** Overview of the proposed system

the operation of an injection pipette using the proposed system.

### 3D Image presentation

The 3D image presentation is based on a previous study [10] which can be used to perform real-time micromanipulation. The process of the 3D image presentation is summarized in the following sections and in Fig. 6.

(a) Capture three images with different focal lengths, and obtain the 3D position and size of a microscopic object from the images.
(b) Transforms pipettes and microscopic objects into the immersive space coordinates.
(c) Present the user with a 3D image in the immersive space.

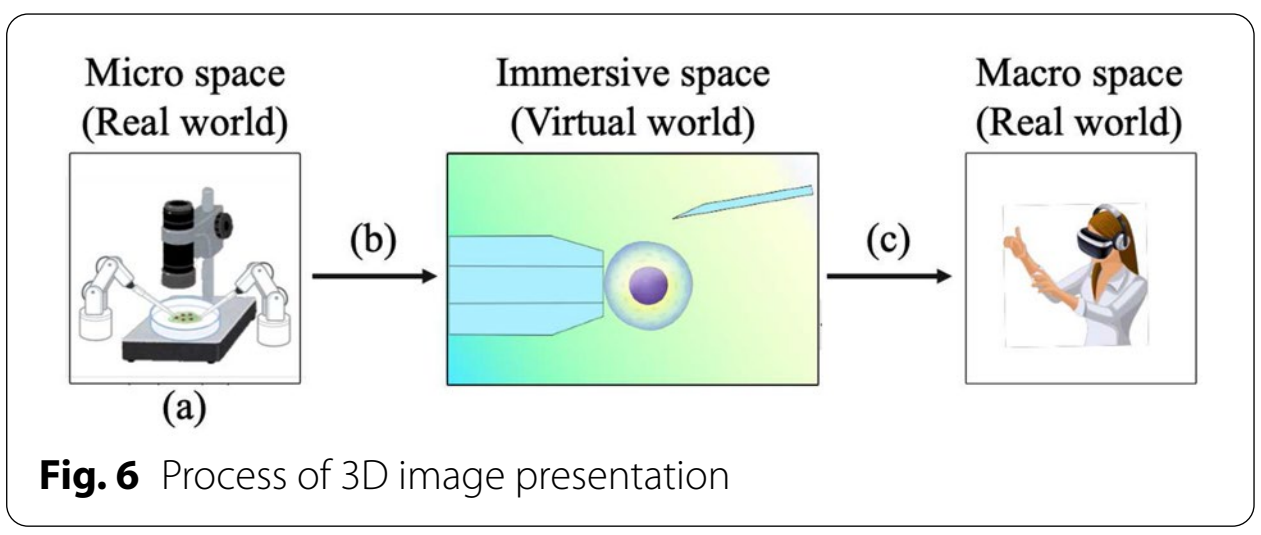

**Fig. 6** Process of 3D image presentation

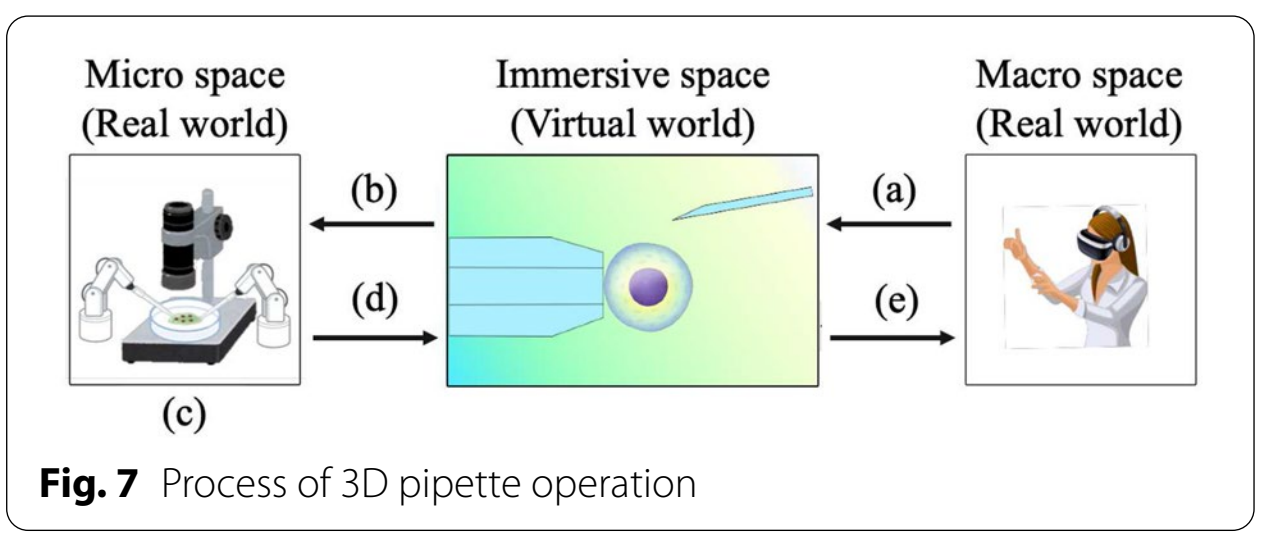

**Fig. 7** Process of 3D pipette operation

The immersive micromanipulation system acquires the positions of the microscopic objects from their images and the manipulator's encoder and places them in an immersive space 1000× larger than the real objects.

### Pipette operation using 3D operation interface

The process of the pipette operation using the 3D operation interface is summarized in the following sections and in Fig. 7.

(a) The user moves a virtual pipette in an immersive space using a 3D operation interface.
(b) Actual pipette moves based on the virtual pipette.
(c) Acquire 3D position of the microscopic object.
(d) In immersive space, virtual microscopic object moves based on actual microscopic object.
(e) Present the user with a 3D microscopic object in the immersive space.

The user is immersed in an immersive space and performs micromanipulation by moving a virtual pipette.

## Algorithm of the immersive micromanipulation system

Algorithm 1 shows the flow of an immersive micromanipulation system. $p$ represents the position vector of the pipette, and $e$ represents the position vector of the embryo to be manipulated. The subscripts $w$, $i$, and $c$ indicate that the position vectors are in the world coordinate system, camera coordinate system, and immersive coordinate system, respectively. The subscript 0 indicates the initial position. $M$ represents the matrices to transform coordinates. Figure 8 shows the image of each coordinate system.

**Algorithm 1** Algorithm of the immersive micromanipulation system

1: $p_{w0} \leftarrow initial\ vector$
2: $p_{i0} \leftarrow M_{wi} p_{w0}$
3: $p_w \leftarrow p_{w0}$
4: $p_i \leftarrow p_{i0}$
5: **loop**
6: Obtain images.
7: $e_c \leftarrow new\ vector$
8: $e_i \leftarrow M_{ci} e_c$
9: User moves the virtual pipettes in the immersive space by 3D operation interface.
10: $p_i \leftarrow new\ vector$
11: $p_w \leftarrow p_{w0} + M_{iw}(p_i - p_{i0})$
12: $\dot{p}_w \leftarrow M_{iw} \dot{p}_i$
13: **end loop**

The immersive space is configured using a real-time 3D imaging microscope [10], which corresponds to lines 1–4 of Algorithm 1. A mapped model of the immersive space configuration is shown in Fig. 9. $P$ and $O$ represent the pipette position and the microscopic object position, respectively. $E$ and $H$ represent the operator's eye position obtained from HMD and the operator's hand position obtained from the 3D operation interface, respectively. $M$ indicates the mapping. The subscripts $w$, $i$, and $c$ indicate the world coordinate system, camera coordinate system, and immersive coordinate system, respectively. The 3D imaging microscope obtains the position of the pipette in immersive space based on the world coordinate of the pipette obtained from the encoder and obtains the position of the microscopic object in the immersive space from the camera coordinate of the microscopic object.

A mapped model of the micromanipulator operation is shown in Fig. 10, which corresponds to lines 6–12 of Algorithm 1. The real-time 3D microscope obtains the position of the microscopic object in the immersive space from the camera coordinate of the microscopic object. The operator inputs in the system the position of the operator's hand through the 3D operation interface. The pipette moves based on the position of the operator's hand in the immersive space. The actual pipette moves based on the position of the pipette in the immersive space. By repeating this process, the operator can freely operate the pipette.

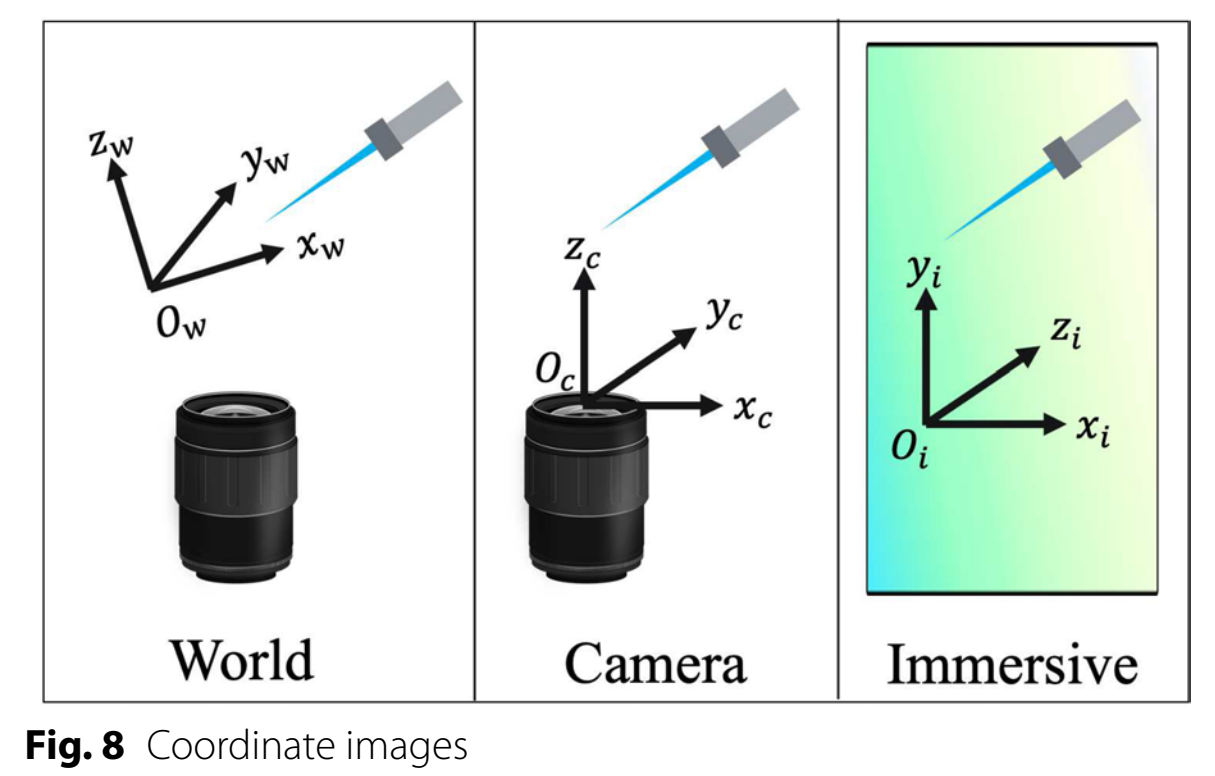

**Fig. 8** Coordinate images

## Evaluation of the operation speed

### Experimental method

We conducted evaluation experiments to verify that the proposed system improved the operation speed of an injection pipette. Figure 11 shows the schematic of this experiment. The subject orbited the injection pipette in the $yz$-plane containing the center of the fixed micro-bead for 15 s. The subjects were instructed to move the injection pipette as quickly as possible. The subjects were asked not to apply the injection pipette to the microbead. If the microbead was dropped, it was considered a failure. However, no subject dropped the microbeads. No limitation was placed on the distance between the injection pipette and the beads. The material of the microbeads was polystyrene, and their diameter was 100 μm. The operation speed is evaluated based on the number of laps. The experiment was conducted under six different conditions, as shown in Table 1, with different image presentation and operation methods. Image presentation conditions included a conventional 2D display and a 3D image presented to the HMD. Operation interface conditions were a conventional joystick, a stationary pen-type interface, and a wearable glove-type interface. In condition (a), the subjects viewed 2D images and used a joystick as the operation interface. In condition (b), the subjects viewed 3D images and used a joystick as the operation interface. In condition (c), the subjects viewed 2D images and used a stationary pen-type interface as the operation interface. In condition (d), the subjects viewed 3D images and used a stationary pen-type interface as the operation interface. In condition (e), the subjects viewed 2D images and used a wearable glove-type interface as the operation interface. In condition (f), the subjects viewed 3D images and used a wearable glove-type interface as the operation interface.

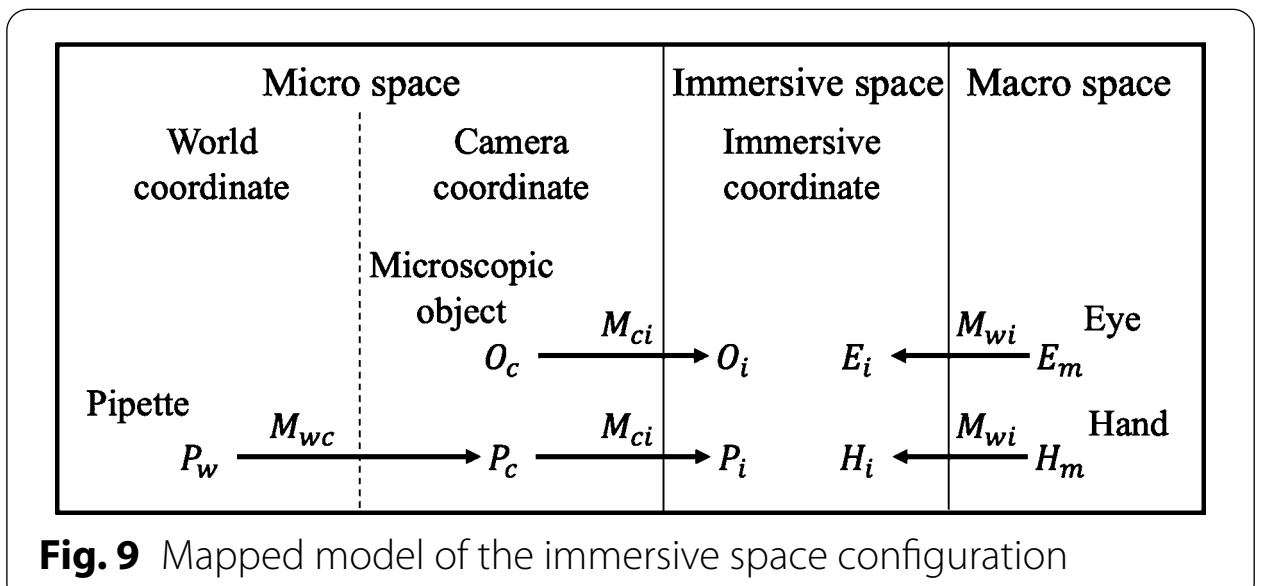

**Fig. 9** Mapped model of the immersive space configuration

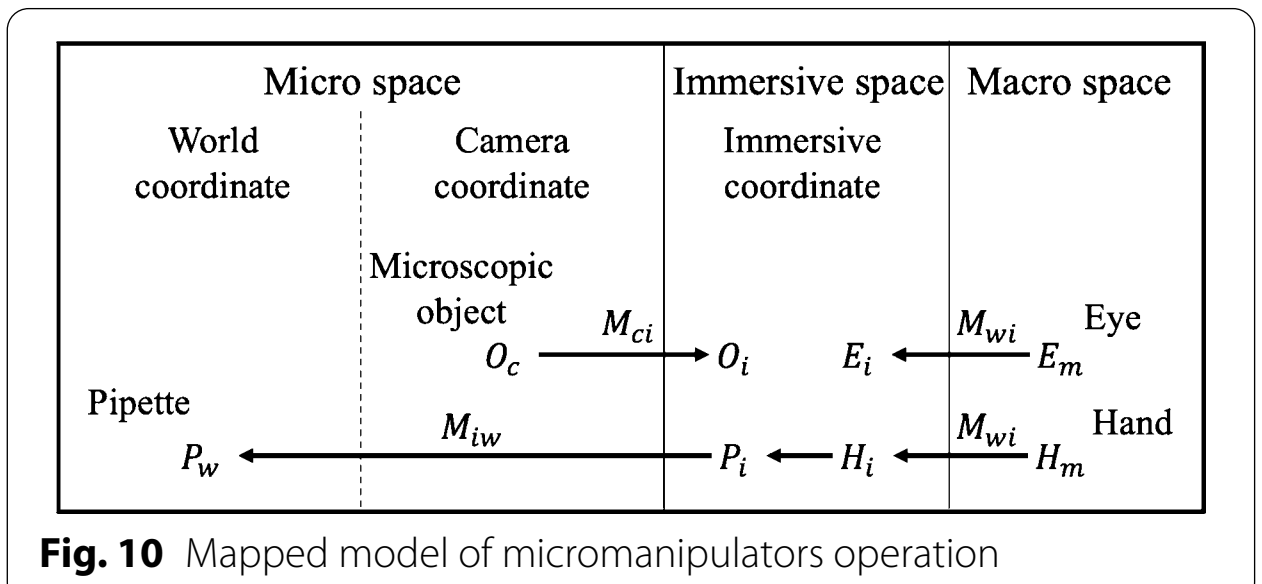

**Fig. 10** Mapped model of micromanipulators operation

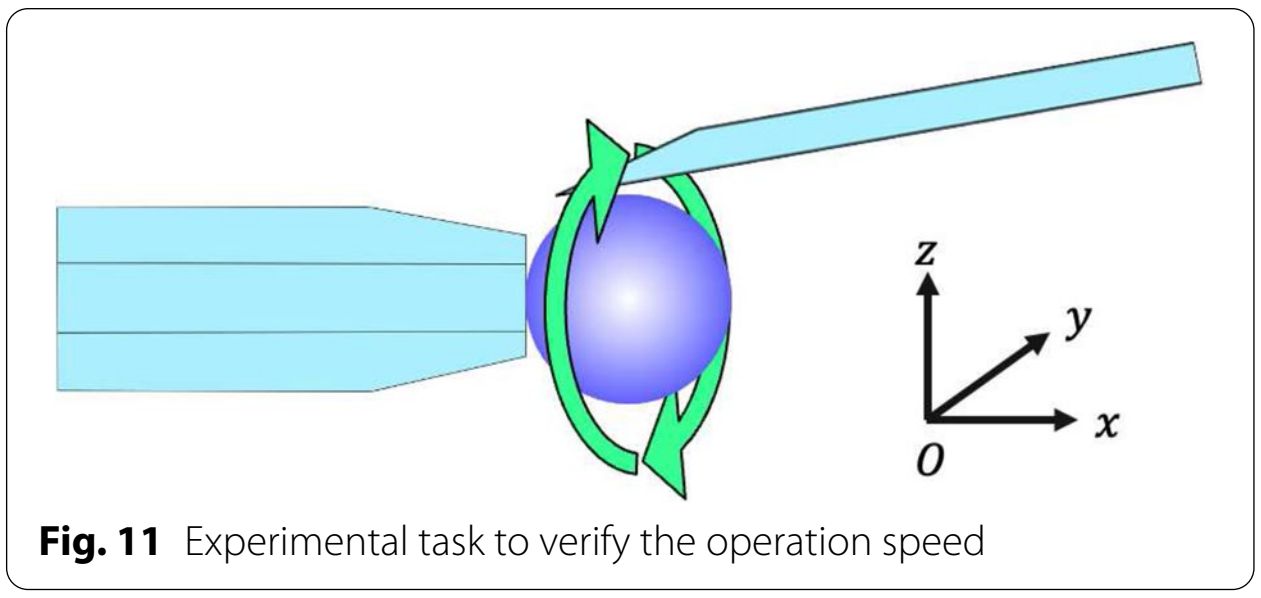

**Fig. 11** Experimental task to verify the operation speed

**Table 1** Experimental conditions

| Condition | Image presentation | Operation interface |
|---|---|---|
| (a) | 2D | Joystick |
| (b) | 3D | Joystick |
| (c) | 2D | Pen-type |
| (d) | 3D | Pen-type |
| (e) | 2D | Glove-type |
| (f) | 3D | Glove-type |

Subjects in each condition are shown in Fig. 12. Figures 13 and 14 show the 2D and 3D images displayed subjects during the experiment, respectively. All five subjects had no experience in micromanipulation and were between 20 and 25 years old. Each subject performed the task five times under all conditions. To make a comparison, a skilled operator engaged in micromanipulation for more than five years also performed the same task using a 2D image and a joystick.

We performed a Friedman test [17], a Kruskal–Wallis test [18], and Dunn's test [19] on the number of laps for each condition with the image presentation method and the operation interface as factors. To investigate the highest speed operation interface in an immersive micromanipulation system, we also performed a Wilcoxon rank-sum test [20] on the number of laps for each subject between conditions (d) and (f).

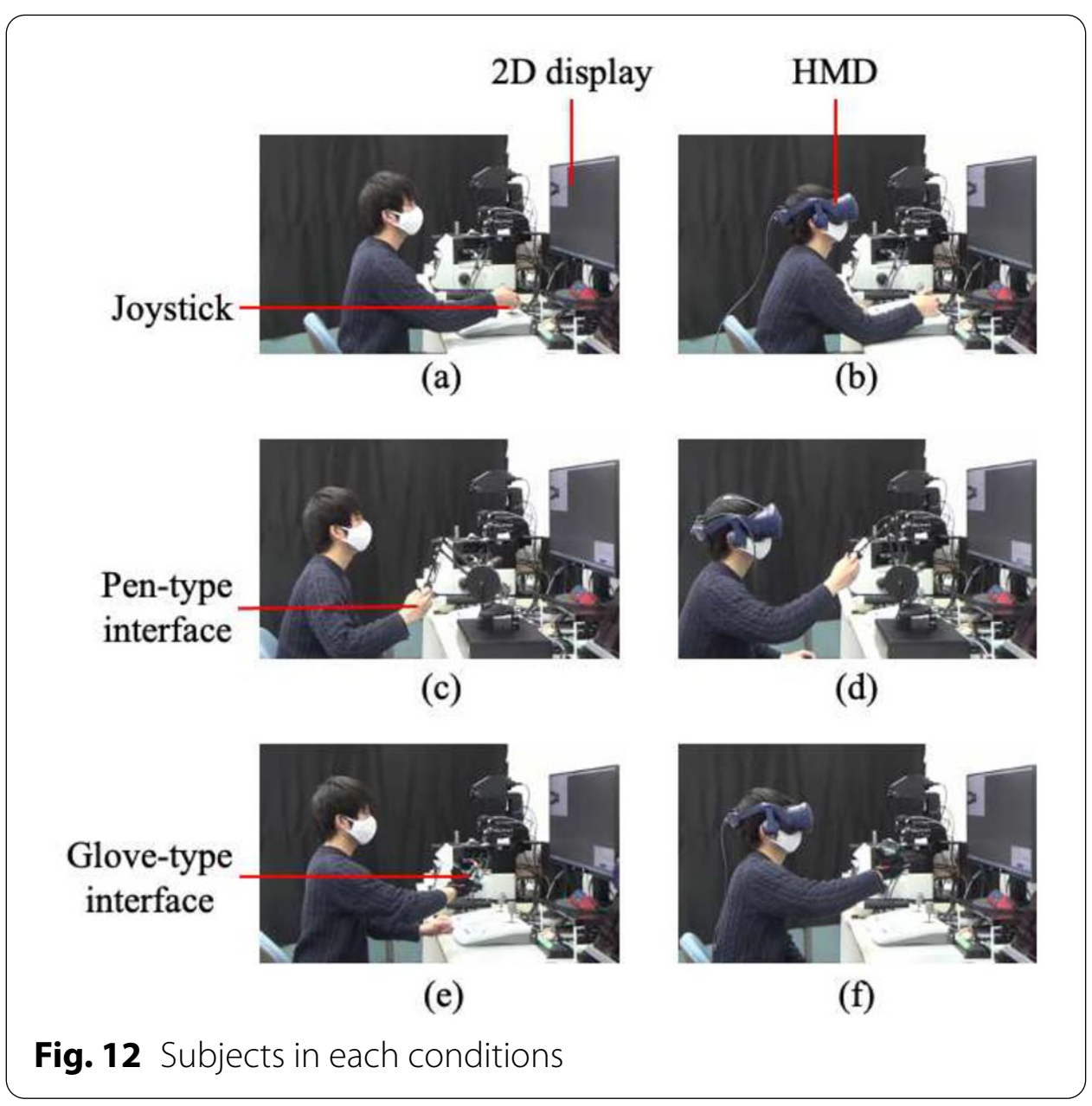

**Fig. 12** Subjects in each conditions

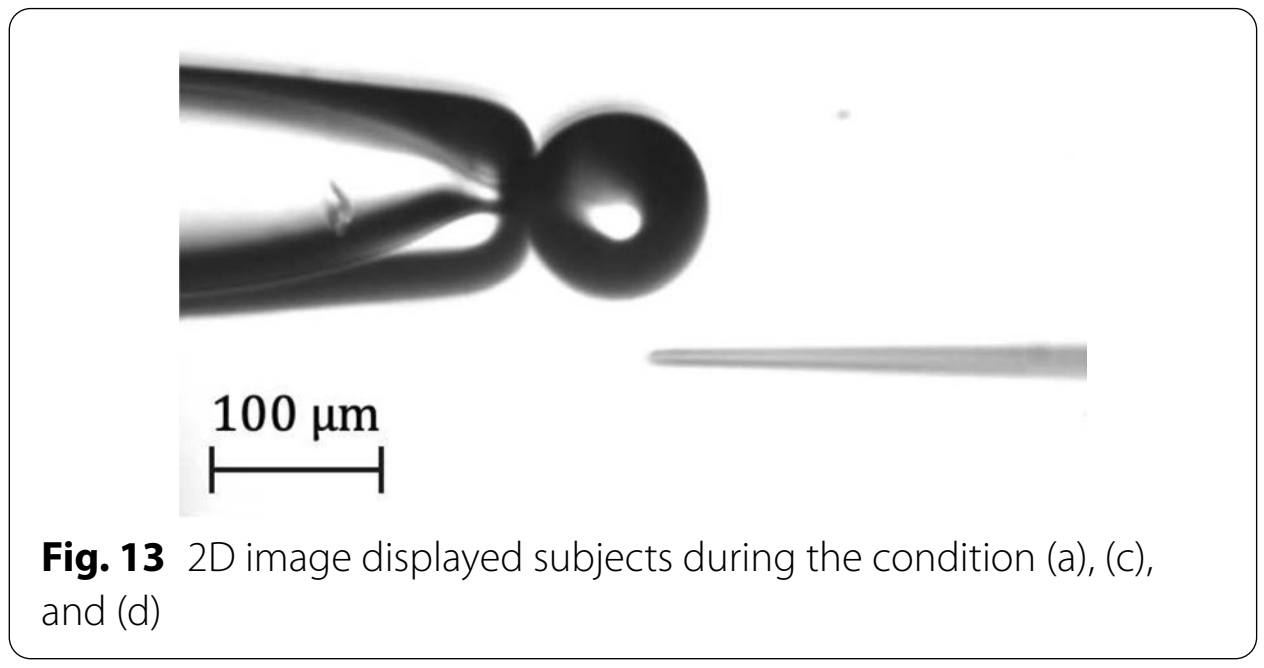

**Fig. 13** 2D image displayed subjects during the condition (a), (c), and (d)

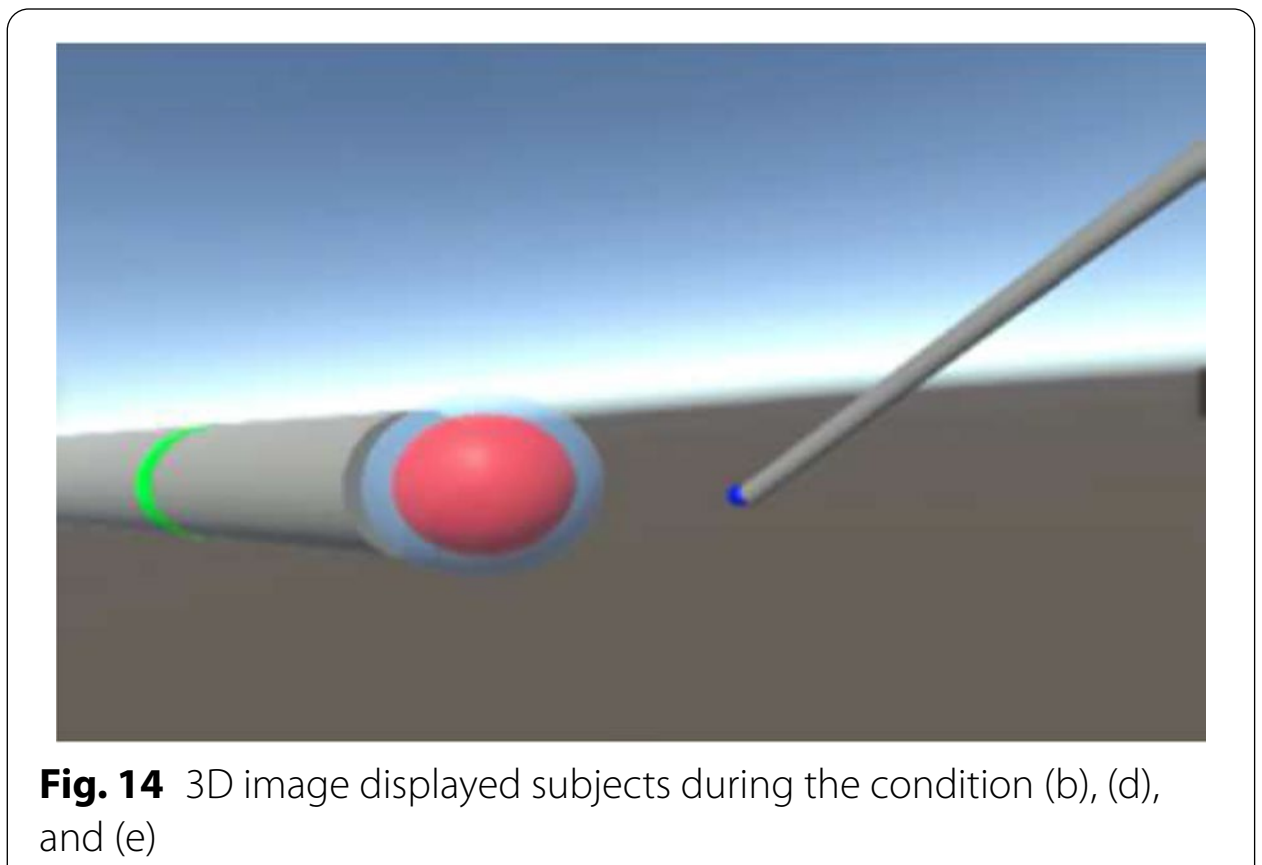

**Fig. 14** 3D image displayed subjects during the condition (b), (d), and (e)

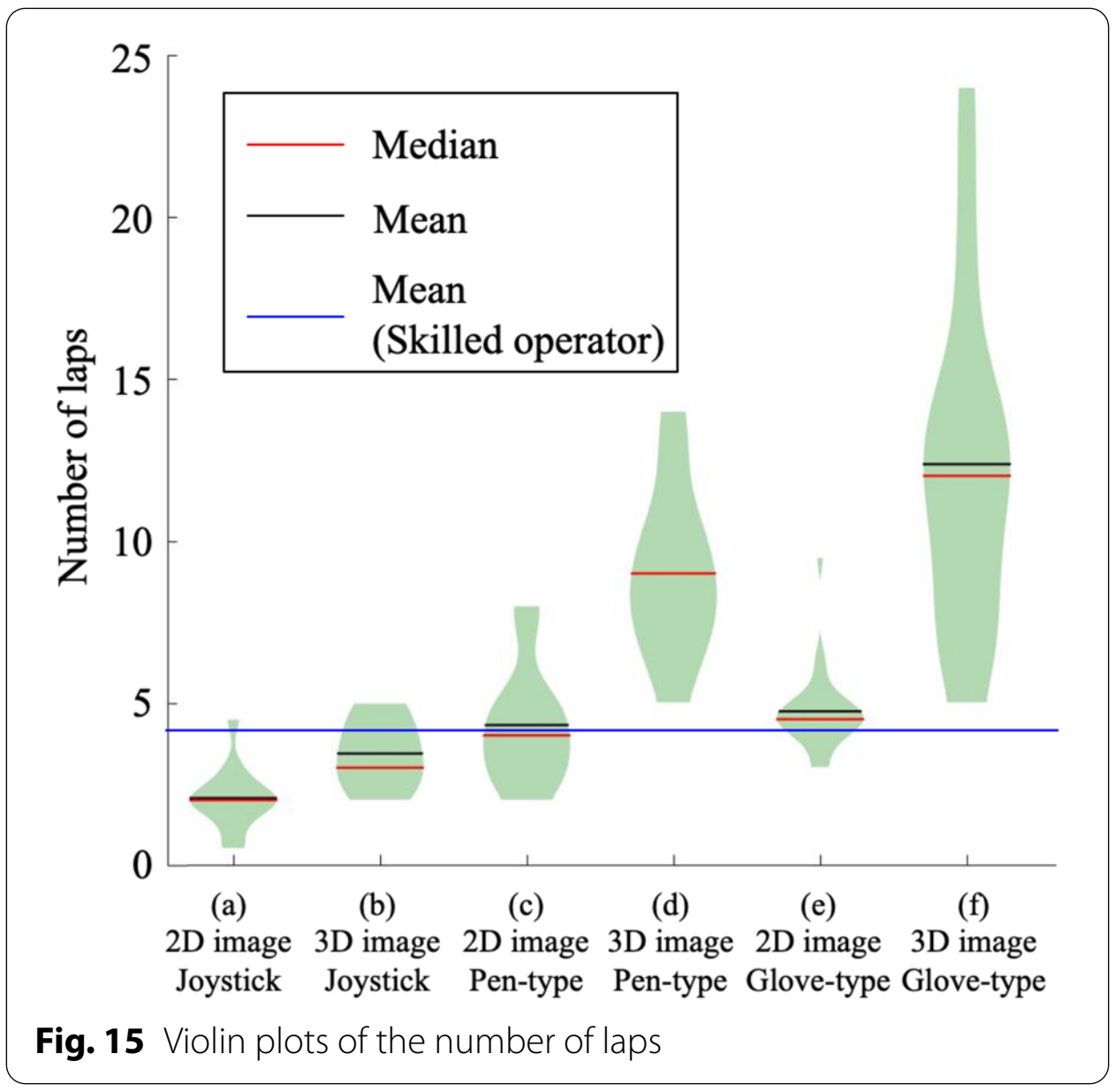

**Fig. 15** Violin plots of the number of laps

### Experimental results

Figure 15 shows the violin plot [21] of the number of laps for each condition. In the violin plot, the red and black lines represent the median and mean values for each condition, respectively. Table 2 and 3 show the average number of laps and the standard deviation of laps for each subject in each condition, respectively. A through E in the tables represent subjects.

The mean laps and standard deviation of laps of the skilled operator were 4.0 and 0.4, respectively. In Fig. 15, the blue line represents the mean laps of the skilled operator. In conditions (d) and (f), the mean laps of the beginner subject were all higher than the skilled operator and the beginner subjects' standard deviation for laps was lower than the skilled operator.

Subject D orbited the injection pipette the most out of all 25 trials in condition (a), the condition using conventional 2D image and joystick operation. The pipette's trajectories of the trials with the highest number of laps for each condition for subject D are shown in Figs. 16 and 17. Figure 16 shows the pipette's trajectories in the *yz*-plane for conditions (a), (c), and (e), 2D image presentation. Figure 17 shows the pipette's trajectories in the *yz*-plane for conditions (b), (d), and (f), 3D image presentation.

We conducted a Friedman test and found that both the image presentation method and the operation interface independently had significant effects on the operation speed ($p < 0.01$). The changes in conditions also significantly affected the change in the laps (Kruskal–Wallis test: $p < 0.01$); therefore, the multiple comparison method, Dunn's test, was used to investigate the best conditions. Conditions (d) and (f) each had significantly more laps than conditions (a), (b), (c), and (e) ($p < 0.01$). There was no significant difference between conditions (d) and (f) ($p > 0.05$). Conditions (c) and (e) each had significantly more laps than condition (a) ($p < 0.01$).

For each subject, the Wilcoxon rank-sum test determined whether the stationary pen-type interface or the wearable glove-type interface was better as an operation interface when using 3D image presentation. For subjects B, C, and D, the number of laps was significantly higher using the wearable glove-type interface than the stationary pen-type interface (B: $p = 0.02$, C: $p = 0.02$, D: $p = 0.01$).

**Table 2** Mean laps

| Condition | Image presentation | Operation interface | A | B | C | D | E | all |
|---|---|---|---|---|---|---|---|---|
| (a) | 2D | Joystick | 1.8 | 2.2 | 2.1 | 3.4 | 0.8 | 2.1 |
| (b) | 3D | Joystick | 3.5 | 3.7 | 2.8 | 4.7 | 2.5 | 3.4 |
| (c) | 2D | Pen-type | 7.2 | 2.9 | 3.8 | 4.3 | 3.4 | 4.3 |
| (d) | 3D | Pen-type | 12.6 | 9.5 | 6.6 | 8.9 | 7.4 | 9.0 |
| (e) | 2D | Glove-type | 4.0 | 4.2 | 4.3 | 5.7 | 5.5 | 4.7 |
| (f) | 3D | Glove-type | 12.2 | 12.7 | 10.1 | 20.4 | 6.4 | 12.4 |

**Table 3** Standard deviation of laps

| Condition | Image presentation | Operation interface | A | B | C | D | E | all |
|---|---|---|---|---|---|---|---|---|
| (a) | 2D | Joystick | 0.2 | 0.2 | 0.2 | 1.0 | 0.4 | 1.0 |
| (b) | 3D | Joystick | 0.9 | 0.7 | 0.7 | 0.4 | 0.4 | 1.0 |
| (c) | 2D | Pen-type | 1.2 | 0.7 | 0.8 | 0.7 | 1.2 | 1.8 |
| (d) | 3D | Pen-type | 1.5 | 1.5 | 0.9 | 0.7 | 1.7 | 2.5 |
| (e) | 2D | Glove-type | 0.5 | 0.7 | 0.5 | 1.9 | 0.7 | 1.2 |
| (f) | 3D | Glove-type | 2.3 | 1.0 | 1.6 | 2.8 | 0.9 | 5.0 |

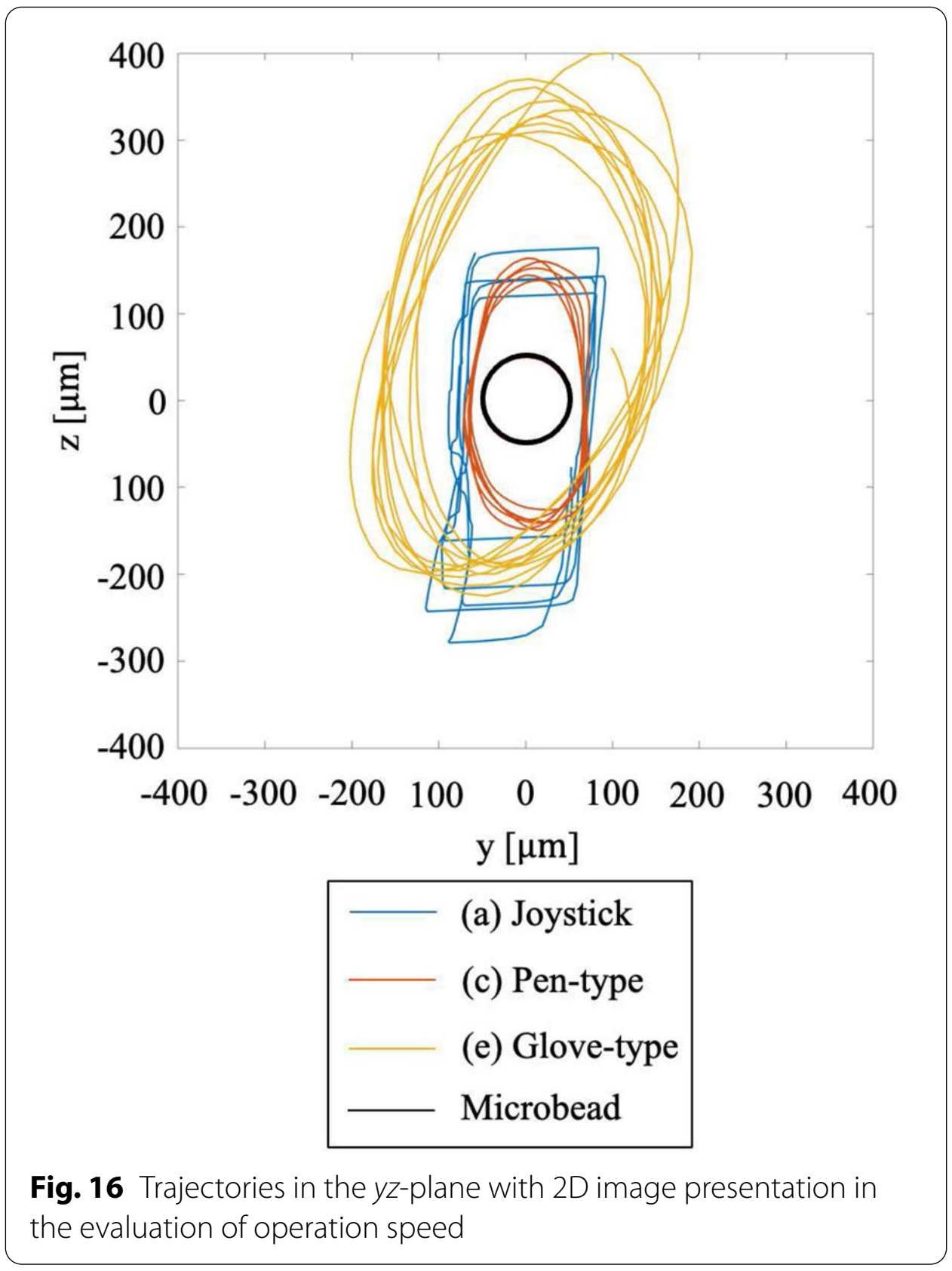

**Fig. 16** Trajectories in the *yz*-plane with 2D image presentation in the evaluation of operation speed

**Fig. 17** Trajectories in the *yz*-plane with 3D image presentation in the evaluation of operation speed

The operation speed of the injection pipette increased when the image presentation method was changed from 2D to 3D and the operation interface was changed from a conventional joystick to a 3D operation interface. Therefore, the immersive micromanipulation system, which combines 3D image presentation and 3D operation interface manipulation, resulted in the highest operation speed. The three subjects had the highest operation speed when the operation interface was a wearable glove-type interface in the immersive micromanipulation system.

## Evaluation of the operation accuracy

### Experimental method

Experiments were conducted to verify that the proposed system improved the operation accuracy of an injection pipette. Figure 18 shows the schematic of this experiment. The subjects touched the center of the fixed microbead with an injection pipette from the $x$-axis. The initial position of the injection pipette tip was 150 to 250 μm in the $x$-axis direction, 100 to 200 μm in the $y$-axis direction, and $-300$ to 300 μm in the $z$-axis direction, as viewed from the center of the microbead. The conditions, subjects, and the number of trials were the same as in evaluation of the operation speed. The operation accuracy was evaluated based on the error of the pipette in the $z$ direction from the center of the microbead at the end of the task. The material of the microbead was polystyrene, and their diameter was 100 μm. Because microbeads cannot be penetrated, the task was considered finished when the tip of the injection pipette hit the microbeads or when the subjects reckoned that they touched the injection pipette tip to the microbeads. Therefore, the microbeads were not excessively pushed during the experiment. In addition, subjects were instructed not to move the injection pipette in the positive *x*-axis direction, that is, away from the microbeads. The error was calculated using the relative positions of the microbead and the injection pipette measured using a real-time 3D imaging microscope [10]. We also measured the time to complete each task.

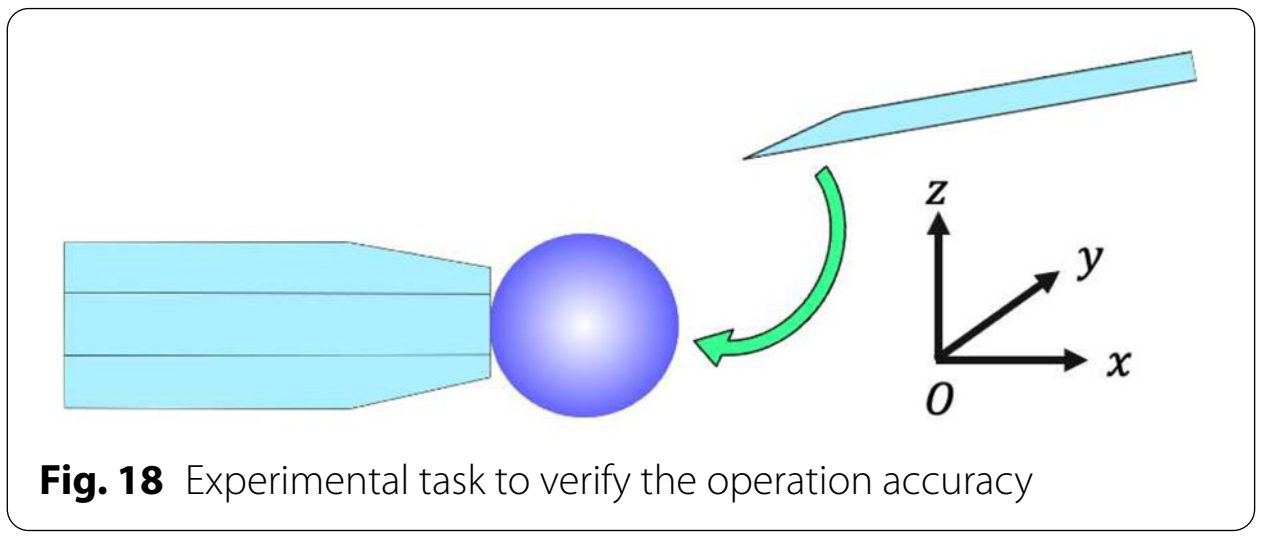

**Fig. 18** Experimental task to verify the operation accuracy

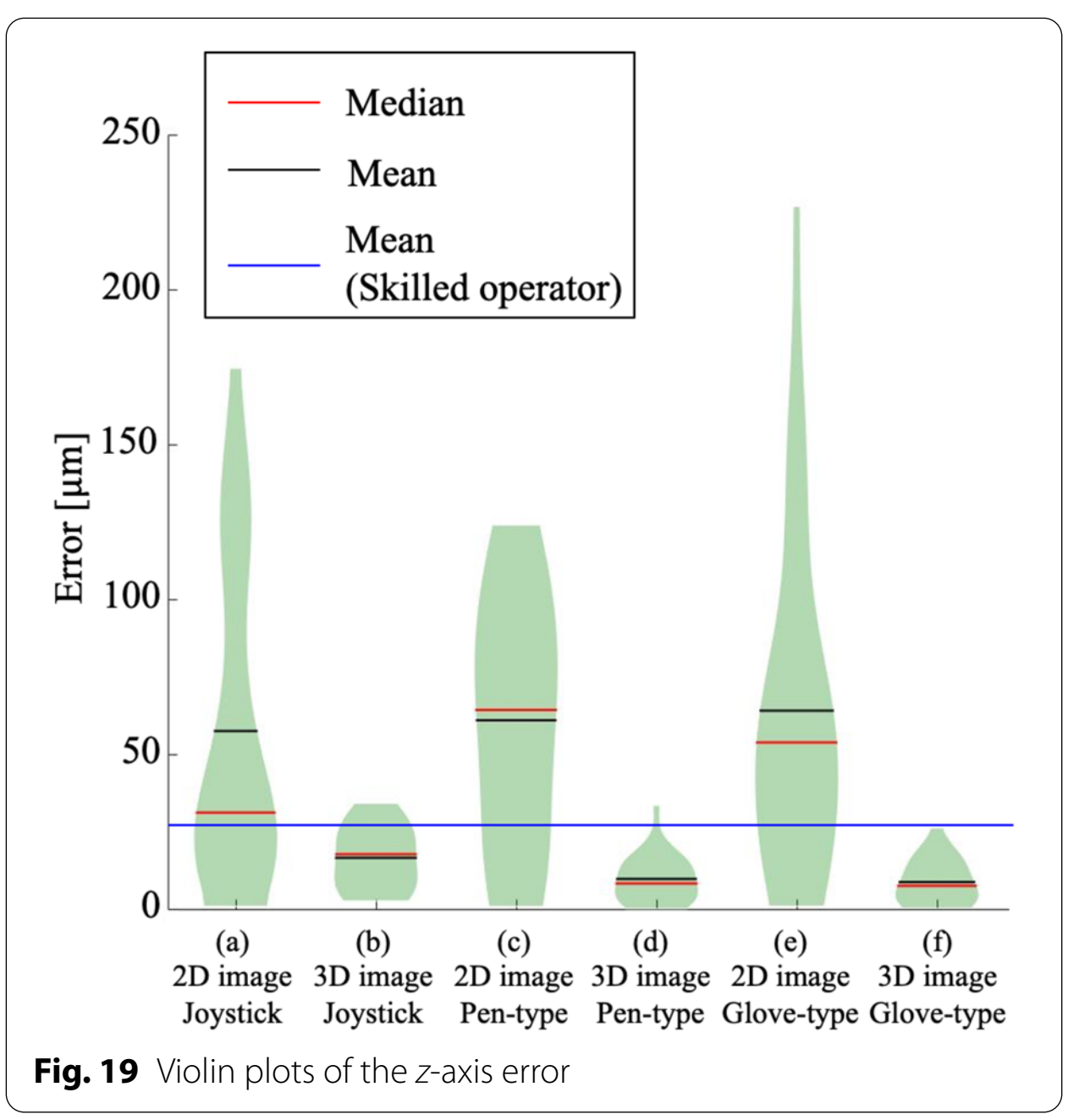

**Fig. 19** Violin plots of the *z*-axis error

We performed a Friedman test, a Kruskal–Wallis test, and Dunn's test on the *z*-axis error and the time for each condition. To investigate the most accurate operation interface in an immersive micromanipulation system, we also performed a Wilcoxon rank-sum test [20] on the *z*-axis error and the time for each subject between conditions (d) and (f).

## Experimental results

Figures 19 and 20 show the violin plot of the *z*-axis error and the time to complete a task for each condition, respectively. In the violin plot, red and black lines are median and mean values for each condition, respectively. Tables 4 and 5 show the mean *z*-axis error and the standard deviation of the *z*-axis error for each subject in each condition, respectively. Table 6 and 7 show the mean time to complete a task and the standard deviation of the time to complete a task for each subject in each condition, respectively. A through E in the tables represent subjects.

The mean error and standard deviation of error of the skilled operator were 27.2 μm and 6.5 μm, respectively. In Fig. 19, the blue line represents the mean error of the skilled operator. In conditions (b), (d), and (f), the mean errors of the beginner subjects were all higher than the skilled operator.

Subject D manipulated the injection pipette with the least error out of all 25 trials in condition (a), the condition using conventional 2D image and joystick operation.

**Table 4** Mean error

| Condition | Image presentation | Operation interface | A | B | C | D | E | all |
|---|---|---|---|---|---|---|---|---|
| (a) | 2D | Joystick | 120.0 μm | 93.4 μm | 15.1 μm | 18.1 μm | 41.0 μm | 57.5 μm |
| (b) | 3D | Joystick | 18.6 μm | 16.9 μm | 12.8 μm | 7.1 μm | 27.5 μm | 16.6 μm |
| (c) | 2D | Pen-type | 72.9 μm | 71.7 μm | 48.3 μm | 53.7 μm | 58.2 μm | 61.0 μm |
| (d) | 3D | Pen-type | 4.0 μm | 11.1 μm | 8.9 μm | 7.1 μm | 17.5 μm | 9.7 μm |
| (e) | 2D | Glove-type | 36.6 μm | 90.0 μm | 67.6 μm | 17.7 μm | 108.4 μm | 64.0 μm |
| (f) | 3D | Glove-type | 10.1 μm | 2.6 μm | 12.8 μm | 8.6 μm | 9.4 μm | 8.7 μm |

**Table 5** Standard deviation of error

| Condition | Image presentation | Operation interface | A | B | C | D | E | all |
|---|---|---|---|---|---|---|---|---|
| (a) | 2D | Joystick | 37.1 μm | 46.4 μm | 10.3 μm | 9.6 μm | 23.7 μm | 51.1 μm |
| (b) | 3D | Joystick | 12.5 μm | 7.0 μm | 6.4 μm | 3.6 μm | 4.7 μm | 10.1 μm |
| (c) | 2D | Pen-type | 33.1 μm | 12.6 μm | 53.9 μm | 43.1 μm | 37.3 μm | 39.7 μm |
| (d) | 3D | Pen-type | 4.7 μm | 5.6 μm | 6.2 μm | 3.4 μm | 9.4 μm | 7.7 μm |
| (e) | 2D | Glove-type | 23.2 μm | 61.6 μm | 37.9 μm | 16.2 μm | 64.0 μm | 56.0 μm |
| (f) | 3D | Glove-type | 6.8 μm | 1.4 μm | 3.9 μm | 6.8 μm | 9.0 μm | 7.0 μm |

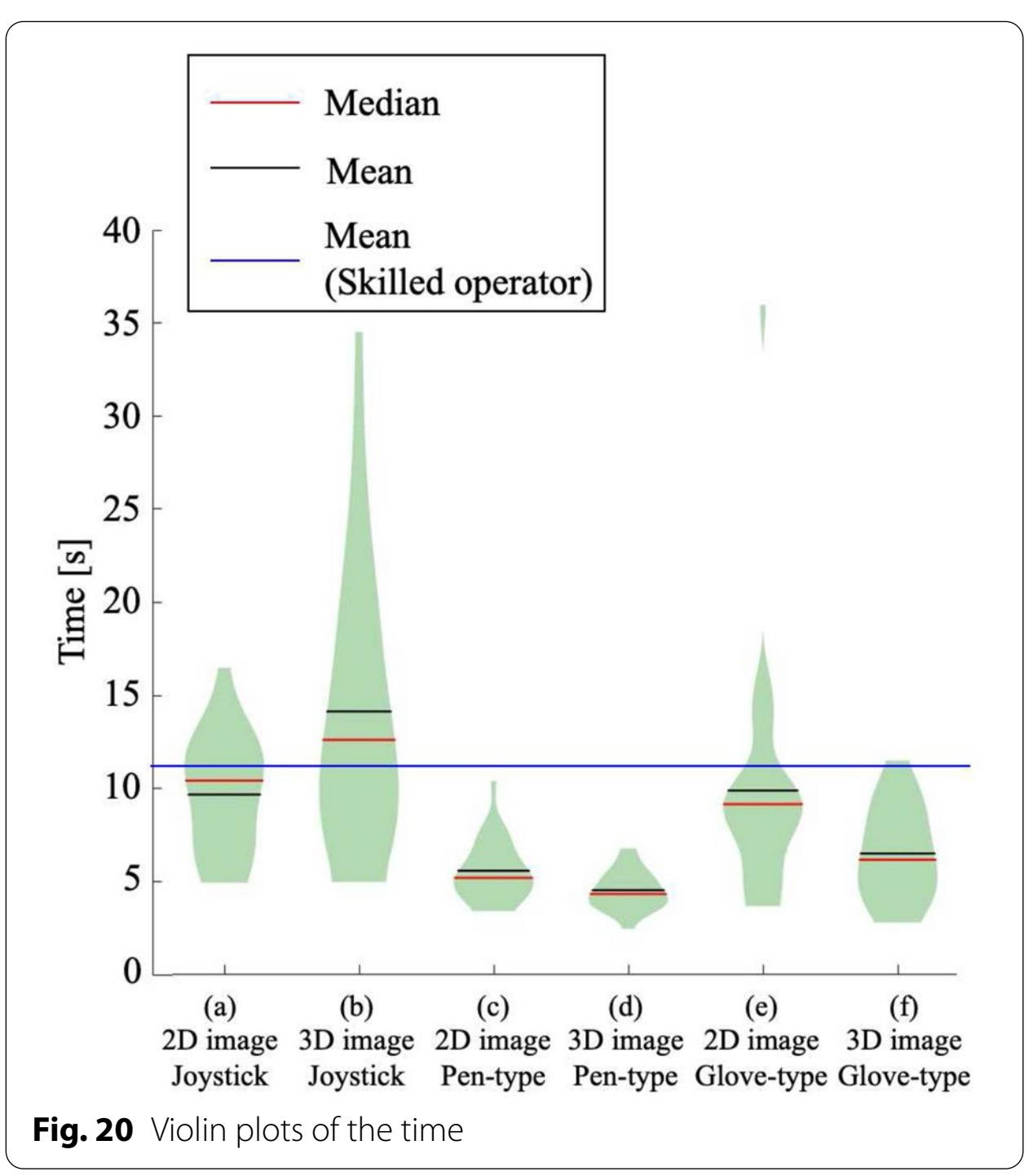

**Fig. 20** Violin plots of the time

The trajectories of the trials with the least error for each condition for subject D are shown in Figs. 21, 22, 23, and 24. Figures 21 and 22 respectively show the pipette's trajectories in the *xz*-plane and *yz*-plane for conditions (a), (c), and (e), 2D image presentation. Figures 23 and 24 respectively show the pipette's trajectories in the *xz*-plane and the *yz*-plane for conditions (b), (d), and (f), 3D image presentation.

We conducted a Friedman test to investigate whether changes in the image presentation methods and the operation interface had independent effects on the operation accuracy. Change in image presentation methods significantly influenced the operation accuracy ($p < 0.01$). However, changes in operation interface had no significant effect on the operation accuracy ($p > 0.05$). The results of the Kruskal–Wallis test represented that the change in conditions affected the operation accuracy ($p < 0.01$). Therefore, the multiple comparison method, Dunn's test, was used to investigate the best conditions. Conditions (d) and (f) each had significantly smaller errors than conditions (a), (c), and (d) ($p < 0.01$). Condition (b) had significantly smaller errors than conditions (c) and (d) ($p = 0.03, p = 0.04$). For each, the Wilcoxon rank-sum test was used to determine whether the stationary pen-type interface or the wearable glove-type interface was better as an operation interface when using 3D image presentation. For subject B, the *z*-axis error was significantly smaller using the wearable glove-type interface than the stationary pen-type interface ($p = 0.03$).

The mean time to complete a task and standard deviation of time to complete a task of the skilled operator were 11.6 s and 3.7 s, respectively. In Fig. 20, the blue line represents the mean time of the skilled operator. In conditions (c), (d), and (f), the mean time to complete

**Table 6** Mean time to complete a task

| Condition | Image presentation | Operation interface | A | B | C | D | E | all |
|---|---|---|---|---|---|---|---|---|
| (a) | 2D | Joystick | 11.2 s | 5.7 s | 8.5 s | 10.3 s | 12.7 s | 9.6 s |
| (b) | 3D | Joystick | 11.5 s | 9.0 s | 8.5 s | 23.3 s | 18.1 s | 14.1 s |
| (c) | 2D | Pen-type | 4.2 s | 5.6 s | 6.3 s | 6.2 s | 5.3 s | 5.5 s |
| (d) | 3D | Pen-type | 3.9 s | 5.1 s | 4.4 s | 3.4 s | 5.7 s | 4.5 s |
| (e) | 2D | Glove-type | 9.1 s | 4.3 s | 10.2 s | 9.1 s | 16.7 s | 9.9 s |
| (f) | 3D | Glove-type | 8.3 s | 4.4 s | 3.9 s | 9.5 s | 6.2 s | 6.5 s |

**Table 7** Standard deviation of time to complete a task

| Condition | Image presentation | Operation interface | A | B | C | D | E | all |
|---|---|---|---|---|---|---|---|---|
| (a) | 2D | Joystick | 0.9 s | 0.8 s | 3.1 s | 1.6 s | 2.3 s | 3.1 s |
| (b) | 3D | Joystick | 2.0 s | 4.3 s | 0.9 s | 6.2 s | 6.3 s | 7.3 s |
| (c) | 2D | Pen-type | 0.8 s | 1.2 s | 2.1 s | 1.5 s | 1.2 s | 1.6 s |
| (d) | 3D | Pen-type | 0.2 s | 0.8 s | 0.8 s | 0.5 s | 0.6 s | 1.0 s |
| (e) | 2D | Glove-type | 2.5 s | 0.4 s | 2.9 s | 1.1 s | 9.9 s | 6.2 s |
| (f) | 3D | Glove-type | 1.2 s | 0.7 s | 0.9 s | 1.3 s | 0.9 s | 2.4 s |

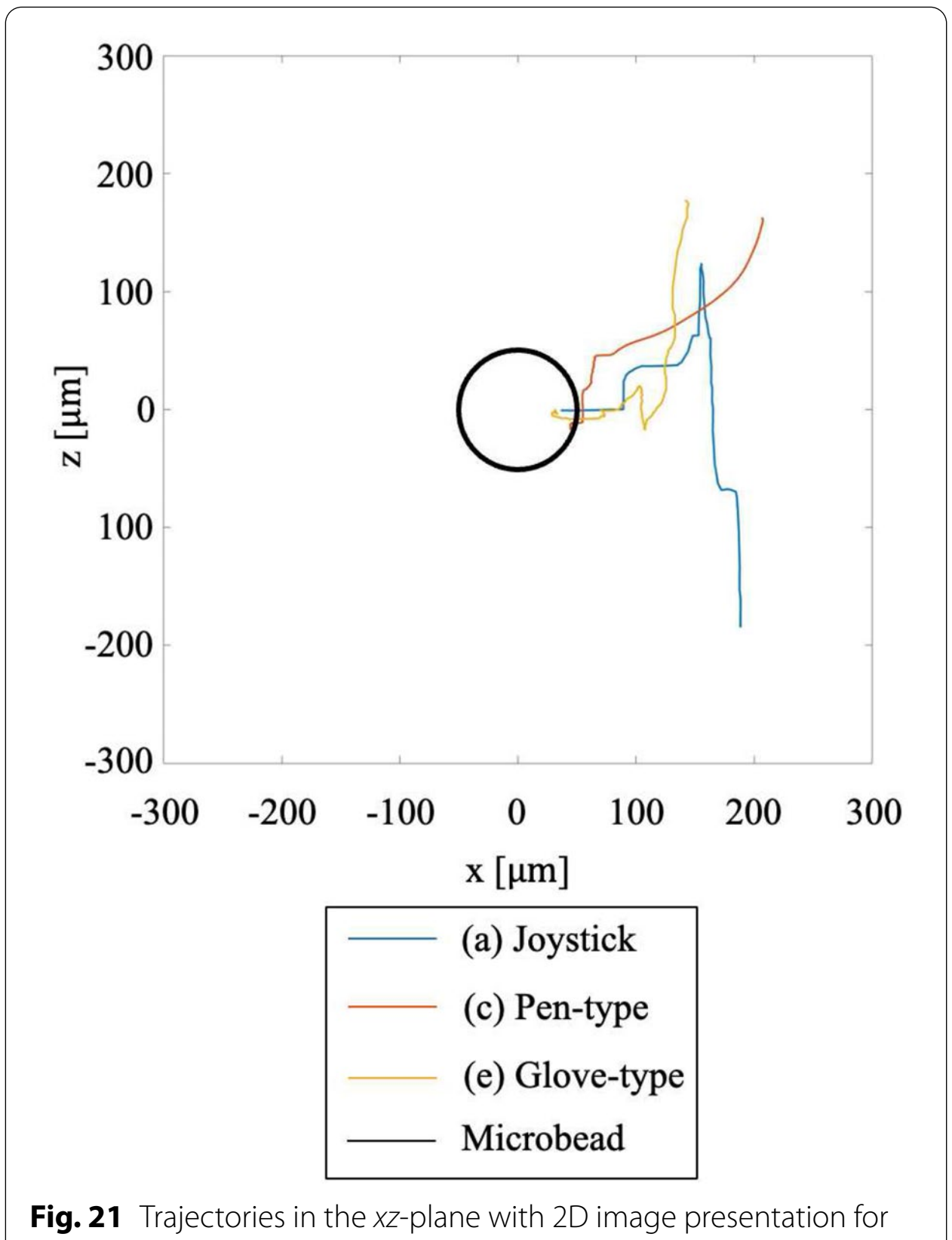

**Fig. 21** Trajectories in the *xz*-plane with 2D image presentation for the evaluation of operation accuracy

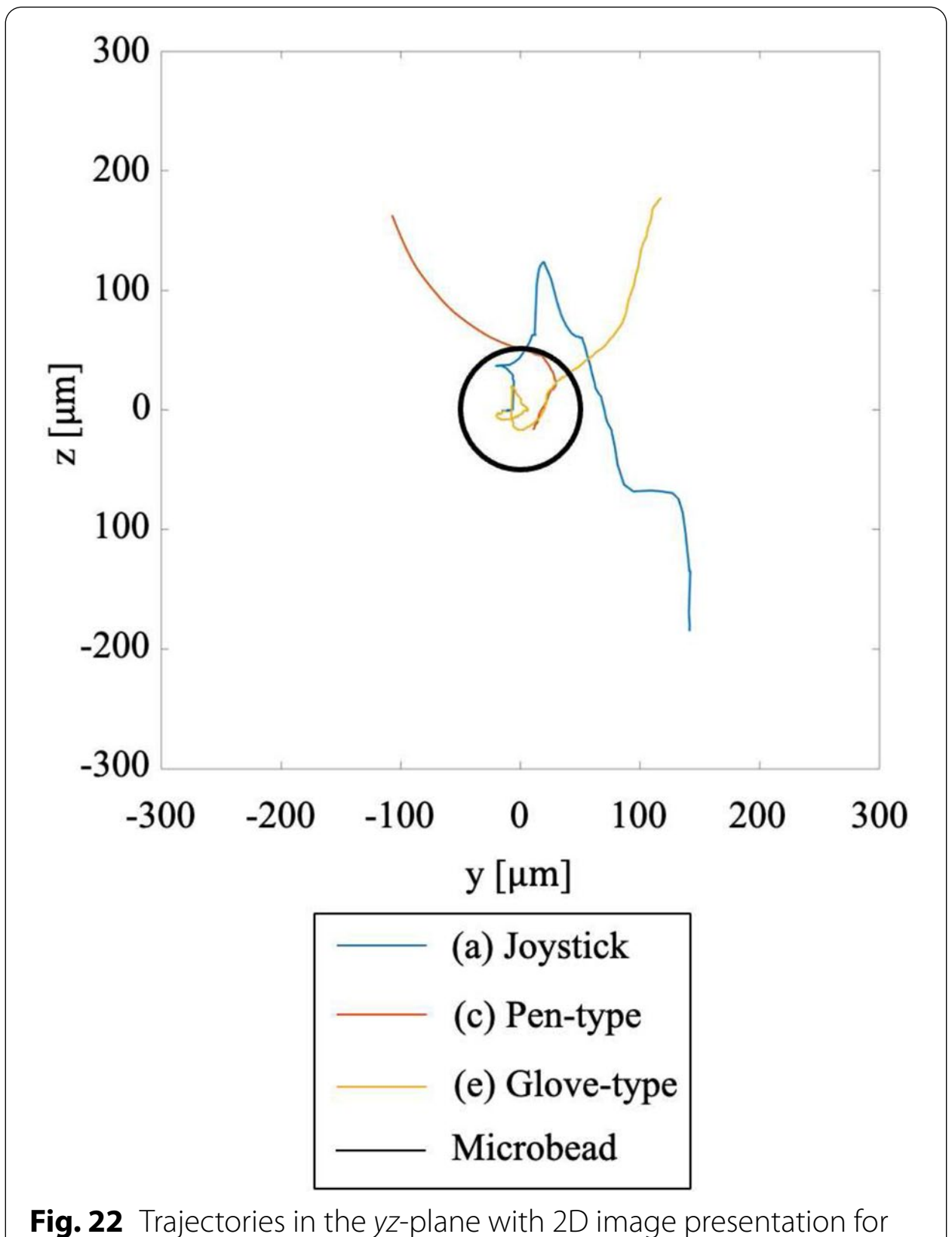

**Fig. 22** Trajectories in the *yz*-plane with 2D image presentation for the evaluation of operation accuracy

a task of the beginner subjects was less than the skilled operator.

We conducted a Friedman test to investigate whether changes in the image presentation methods and the operation interface had independent effects on the time to complete a task. Change of the operation interface significantly influenced the time to complete a task ($p < 0.01$). However, changes in image presentation methods had no significant effect on the time to complete a task ($p > 0.05$). The results of the Kruskal–Wallis test represented that the change in conditions affected the time to complete a task ($p < 0.01$). The multiple comparison method, Dunn's test, was used to investigate the best conditions. Conditions (c) and (d) each had significantly less time to complete a task than (a) and (b) ($p < 0.01$). There was a significant difference in time to complete a task between conditions (c) and (e) ($p = 0.01$). There was a significant difference in time to complete a task between conditions (d) and (e) ($p < 0.01$). Condition (f) had significantly less the time to complete a task than (a) ($p = 0.03$). Condition (f) also had significantly less the time to complete a task than (b) ($p < 0.01$).

For each subject, the Wilcoxon rank-sum test determined whether the stationary pen-type interface or the wearable glove-type interface was better as an operation interface when using 3D image presentation. For subjects A and B, the time to complete a task was significantly less using the stationary pen-type interface than the wearable glove-type interface ($p = 0.01$).

The operation accuracy of the injection pipette was increased by changing the image presentation method from 2D to 3D. In particular, the immersive micromanipulation system that combines 3D image presentation and 3D operation interface resulted in the highest operation accuracy. A subject had the highest operation accuracy when the operation interface was a wearable glove-type interface in the immersive micromanipulation system. We have found that immersive micromanipulation systems enable higher speed operation than conventional systems for tasks that require accuracy. Two subjects had the highest speed operation when the operation interface was a stationary pen-type interface in the immersive micromanipulation system for tasks that require accuracy.

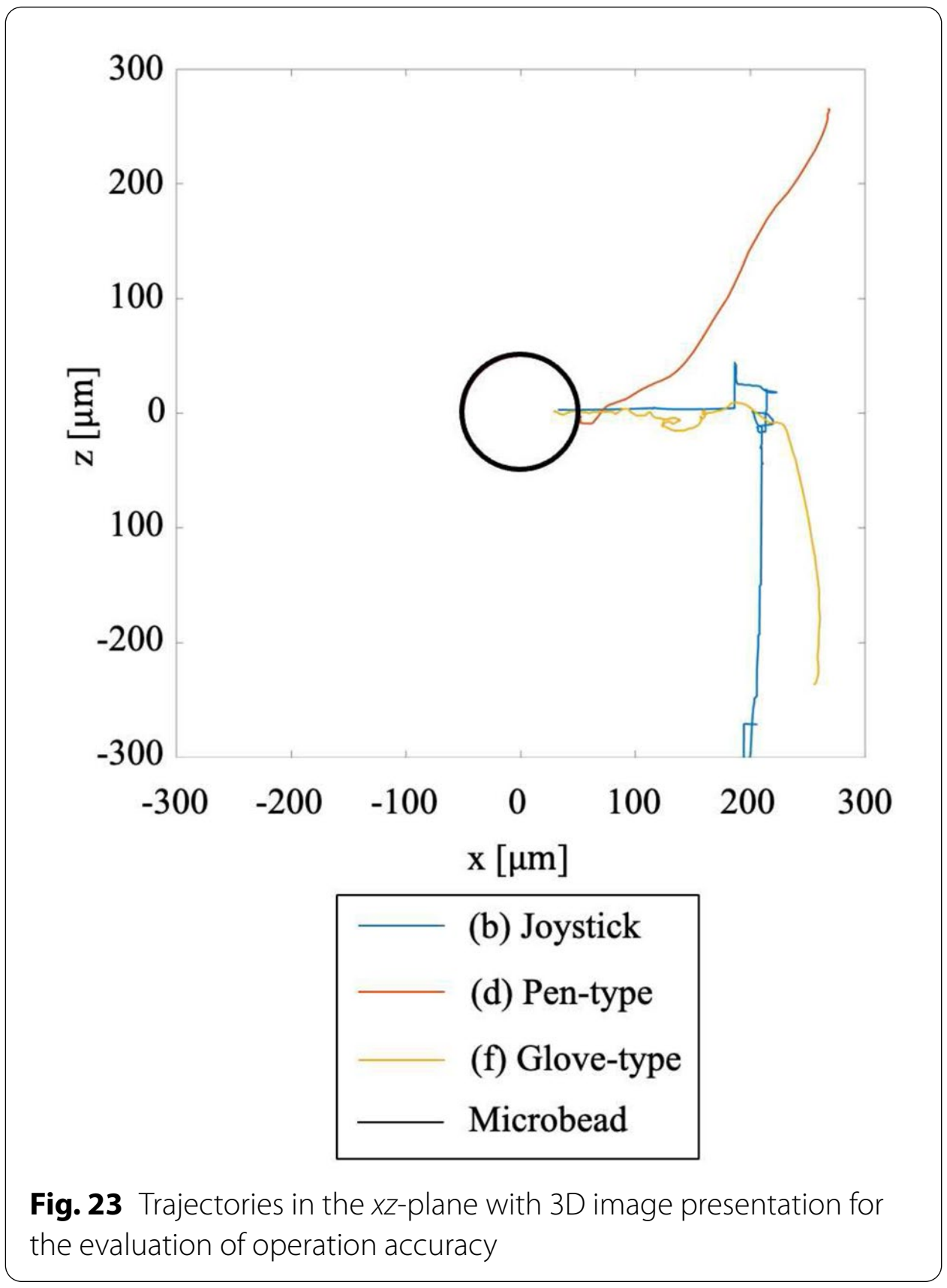

**Fig. 23** Trajectories in the *xz*-plane with 3D image presentation for the evaluation of operation accuracy

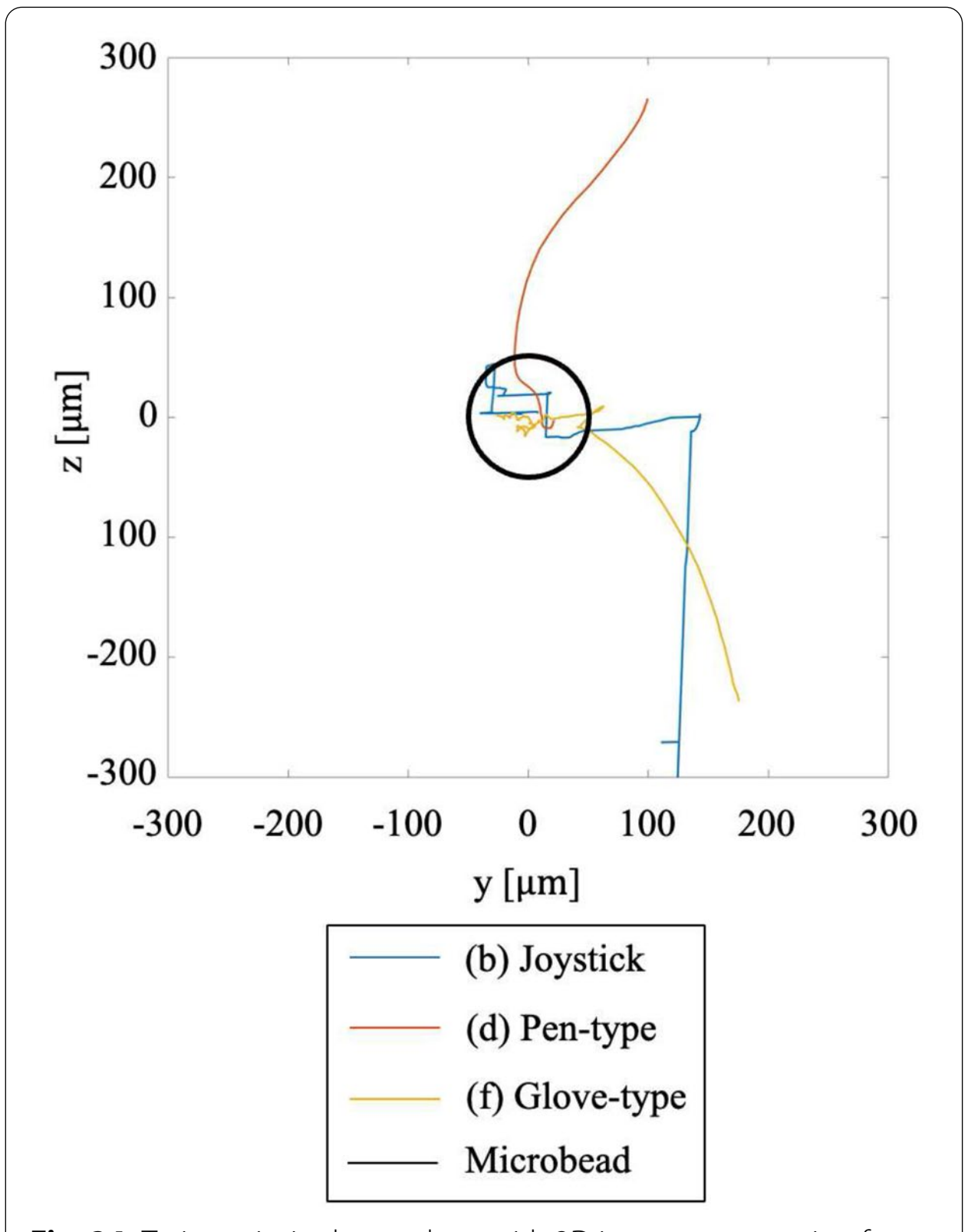

**Fig. 24** Trajectoriesin the *yz*-plane with 3D image presentation for the evaluation of operation accuracy

## Discussion

The 3D image presentation significantly improves both pipette operation speed and accuracy. The 3D image presentation improves the user's depth visibility compared with conventional 2D images. The improved depth visibility allows the user to precisely identify the position of the pipette thus improving the accuracy of pipette operation. In addition, the improved depth visibility allows the user to view the accurate pipette position. Therefore, the operator can confidently manipulate the pipette, which may have resulted in higher operation speeds.

The 3D operation interfaces, including a wearable glove-type interface and a stationary pen-type interface, increase the speed of injection pipette operation. However, the 3D operation interfaces did not improve the accuracy of operation independently. Conventional joystick operation is a combination of 2-DoFs of translation and 1-DoF of rotation. Therefore, the operator cannot move the pipette in all 3-DoFs simultaneously. In contrast, the 3D operation interfaces allow the operator to move the injection pipette in 3-DoFs simultaneously. Therefore, as the operator could move the pipette in 3-DoFs simultaneously, it may have increased the speed of the operation. However, the 3D operation interface cannot move only one axis while the other two axes of the pipette are fixed. The accuracy did not increase because while aligning one axis of the pipette, the other two axes were displaced.

When the operator uses both the 3D image presentation and 3D manipulation interface simultaneously, that is, when using an immersive micromanipulation system, pipette operation speed and accuracy are further improved. The immersive micromanipulation system allows beginners in micromanipulation to operate pipettes with the same speed and accuracy as experienced operators. Because the immersive micromanipulation system is 3D for both visual presentation and pipette operation, the operator is immersed in the micro-space and can operate the pipette as if it were in macro space, which may have further improved the accuracy and speed.

When the image presentation is 2D and the operation interface is 3D, both speed and accuracy remained the same as compared with when the image presentation is 2D and the operation interface is 2D. Although the operator is operating the pipette in 3D, the operator receives 2D images, which makes it difficult to correct the trajectory of the pipette. Therefore, the operator cannot perform accurate micromanipulation. Inaccurate pipetting causes unnecessary pipette movement resulting in speed of operation being reduced.

The overall subject results show no significant difference between the stationary pen-type and the wearable glove-type interfaces when using the immersive micromanipulation system. However, there were significant differences in a few subjects between the stationary pen-type interface and the wearable glove-type interface when using the immersive micromanipulation system. For three subjects, the operation speed with the wearable glove-type interface was significantly higher than that with the stationary pen-type interface. For a subject, the operation accuracy with the wearable glove-type interface was significantly higher than that with the stationary pen-type interface. The wearable glove-type interface provides a more immersive experience for the operator than the stationary pen-type interface. The higher level of immersion would allow the operator to have a perception similar to being in macro-space; this would increase the speed and accuracy of operation. However, for two subjects, the operation speed with the stationary pen-type interface was significantly higher than that with the wearable glove-type interface when performing accurate operations. The stationary pen-type interface provides resistance to the operator, which allows the operator to easily fix the pipette position, as opposed to the wearable glove-type interface, which provides no resistance during operation. Therefore, the pen-type interface may improve the speed and provide accurate operations for some operators. The working space of the stationary pen-type interface used in this experiment has a 191-mm length, 381-mm width, and 267-mm height and the glove-type interface used in this experiment is 5000 mm in length, 5000 mm in width, and 2000 mm in height. There is no significant difference in operation speed and accuracy between the stationary pen-type interface and the wearable glove-type interface; however, the sizes of the workspace are significantly different, so the glove-type interface is recommended when using an immersive micromanipulation system.

## Conclusion

We proposed an immersive micromanipulation system that enables high-speed and high-precision micromanipulation. The operator immersed in the virtual micro-space operates the virtual pipette using a 3D operation interface. The actual pipette is operated based on the position of the virtual pipette. The operation speed and accuracy of the pipette in an immersive micromanipulation system were significantly improved as compared with that of conventional micromanipulation and this was verified by conducting two experiments. When beginners used the immersive fine manipulation system, both operation accuracy and speed were equal to or better than skilled operators' micromanipulation. Future work will include reducing the physical burden on the operator and expanding the 3D manipulation space.

## Supplementary Information

The online version contains supplementary material available at https://doi.org/10.1186/s40648-022-00228-6.

**Additional file 1.** Immersive micromanipulation system.

**Acknowledgements**

Not applicable.

**Author contributions**

KY conducted the experiments and drafted the manuscript. TA conceptualized the work. KY and TF developed the proposed system. TA and MT revised the manuscript. YH supervised the study. All authors read and approved the final manuscript.

**Funding**

This work was partly supported by JST PRESTO (Grant No. JPMJPR18J1), the Murata Science Foundation, and JSPS KAKENHI Grant Number JP22H03630.

**Availability of data and materials**

The datasets used and/or analyzed during the current study are available from the corresponding author on reasonable request.

## Declarations

**Ethics approval and consent to participate**

The protocol of this study was approved by the Ethics Committee of the Graduate School of Engineering, Nagoya University (Approved Number: 20-23).

**Competing interests**

The authors declare that they have no competing interests.

## Publisher's Note

Springer Nature remains neutral with regard to jurisdictional claims in published maps and institutional affiliations.